# Studying Buildings Outlines to Assess and Predict Energy Performance in Buildings: A Probabilistic Approach

Zohreh Shaghaghina, zohreh-sh@tamu.edu
*Texas A&M University, United States*

Fatemeh Shahsavari, abanshahsavar@gmail.com
*Perkins+Will, United States*

Elham Delzendeh, Elham.Delzendeh@bcu.ac.uk
*Birmingham City University, United Kingdom*

**Abstract**
Building energy performance is commonly calculated during the last phases of design, where most design specifications get fixed and are unlikely to be majorly modified based on design programs. Predictive models could play a significant role in informing architects and designers of the impact of their design decisions on energy consumption in buildings during early design stages. A building outline is a significant predictor of the final energy consumption and is conceptually determined by architects in the early design phases. This paper evaluates the impact of a building's outline on energy consumption using synthetic data to achieve appropriate predictive models in estimating a building's energy consumption. Four office outlines are selected in this study, including square, T-, U-, and L-shapes. Besides the shape parameter, other building features commonly used in literature (i.e., Window to Wall Ratio (WWR), external wall material properties, glazing U-value, windows' shading-depth, and building orientation) are utilized in generating data distribution with a probabilistic approach. The results show that buildings with square shapes, in general, are more energy-efficient compared to buildings with T-, U-, and L-shapes of the same volume. Also, T-, U-, and L-shape samples with the same area show very similar behaviors in terms of energy consumption, regardless of their wall-to-floor ratios. Principal Component Analysis (PCA) is applied to assess the variables' correlations on data distribution; the results show that wall material specifications explain about 40% of data variation. Finally, we applied polynomial regression models with different degrees of complexity to predict the synthesized building models' energy consumptions based on their outlines. The results show that degree 2 polynomial models, fitting the data over 98% R squared (coefficient of determination), could be used to predict new samples with high accuracy.

**Keywords**
Building Energy Performance, Building Outline, Wall-to-Floor Ratio, Machine Learning

## 1  Introduction and Background

Buildings account for about 40% of carbon emissions in the US due to their large amount of energy consumption (US Dept. of Energy, 2012). Schematic-phase design decisions could have a huge impact on the buildings' energy use; however, energy performance evaluations are usually postponed to the final phases when design modifications are costly if possible (Hygh *et al.* 2012, Krygiel and Nies 2008). Although schematic design accounts for a relatively low portion of the total design fee, it is responsible for the main environmental impact and operating cost of the final buildings' performance (Krygiel and Nies 2008). Selecting efficient building parameters in the early stages of design could improve buildings' energy performances (Hatem and Karram 2020). Multiple buildings' parameters are studied in literature as effective variables that could have significant impacts on the final energy consumption of buildings, including building's geometry, envelope property, window property, Window to Wall Ratio (WWR), and building's orientation, which are chosen by architects mostly in the early phases. Conceptual buildings' shapes and geometries are one of the important decisions architects take in the early phases, upon which, they further improve the design process (Okudan and Tauhid 2008). Energy simulation tools provide a fairly accurate prediction of a building's energy performance when all specifications are considered in the simulation process; however, many design decisions may not be deterministically made in the early phases, yielding to a poor prediction result (Amasyali and El-Gohary 2018). Achieving efficient design parameters requires investigating multiple design alternatives. However, applying simulation tools for all design alternatives is costly and time-consuming (Shaghaghian and Yan 2020). Also, most energy simulation tools require sufficient knowledge and expertise to implement. Hence, the significance of practical tools in providing performance-based information to the designers in the early stages of design is crucial (Hygh *et al.* 2012).

Data-driven approaches and predictive models have been studied in literature as statistical methods to approximately predict buildings' energy performances without a step-by-step calculations (Seyrfar *et al.* 2021, Amasyali and El-Gohary 2018). Such methodologies are specifically useful for providing performance-based information at a fast pace in the early stages of design to help architects select building parameters based on their target variables (Abarghooie *et al.* 2021, Nourkojouri *et al.* 2021). Although having the information of the existing buildings as the input data for the data-driven approach is undeniably beneficial, the accessibility and difficulty in the information retrieval have propelled many studies into utilizing synthetic models and simulations in providing the required training data for predictive models (Amasyali and El-Gozhary 2018, Reddy 2006). Multiple studies in literature have explored building shape as a key parameter in building energy simulation and predictive models (Asl *et al.* 2016, Hatem and Karram 2020, Nazari and Yan 2021). Rahmani asl (2016) has developed a form-based tool to seamlessly predict a building's energy simulation in the early phases of design. In his study, most form-related parameters, such as building length, width, height, and WWR are considered, while other performance-based parameters such as building orientation wall and window material properties have not been explored (Asl *et al.* 2016). In another study, researchers have investigated buildings shapes along with other performance-based variables to evaluate building energy consumption; however, no predictive model is developed in their study (Hatem and Karram 2020). Also, both studies have used deterministic values in generating their data samples while uncertainty is one of the major issues needed to be considered in existing versus simulated data that could help in yielding more practical results (Shahsavari *et al.* 2019).

The focus of this study is developing regression models to predict building energy performance in the early phases of design considering different buildings' outlines. Four building outlines, i.e., square-, T-, U-, and L-shapes, are selected for this study inspired from a similar study done by Hatem et al., 2020. The present study illustrates the impact of building shape on developing predictive energy models with high accuracy. To generate synthetic data for training the predictive models, we considered eight performance-based building features commonly introduced in literature, namely, building orientation, Window to Wall Ratio (WWR), windows' shading-depth, glazing U-value, wall thickness, wall conductivity, wall density, and wall Specific Heat Capacity (SHC). The last four features are the properties of buildings' wall material used in similar studies (Kouhirostamkolaei *et al.* 2021). Literature shows that, in practice, an existing building's features often deviate from the pre-assumed/simulated values (Reddy 2006). This deviation leads to unexpected errors in estimating the final building's energy consumption (Rezaee *et al.* 2018). However, to avoid such uncertainties, one can perform simulations with artificially added Gaussian noise to the expected values of a building's features and, therefore, estimate the energy consumption along with the possible practical deviations (Macdonald 2002). This paper has conducted an uncertainty approach used in similar studies (Shahsavari and Shaghaghian 2021) to consider the features' deviations reported by literature among existing building and simulation models. In this paper, thermal load and total energy consumption are used interchangeably, referring to a building total energy demand (heating and cooling) to keep occupants in thermal comfort zone. We analysed the data using visualization methods in displaying the data distribution. Principal Component Analysis (PCA) is applied as a feature selection technique to observe the impact of the independent variables used in this study on data variation. Finally, regression models are developed to predict data samples based on their shape outlines and the eight building features implemented in the simulations.

## 2 Research Methodology

This study applies Rhino/Grasshopper, a parametric modelling platform, to generate hypothetical office buildings, with four shape outlines and eight numerical building features commonly used as variables for building energy consumption in literature. A probabilistic approach is utilized in generating the data using the Gaussian distribution. The energy consumptions of the synthesized samples are computed through parametric energy simulation plugins, Ladybug/Honeybee. Data visualization methods are utilized to analyse data distribution. Variables' covariances have been analysed through PCA as a feature selection method, and finally, polynomial regression models are applied to get the best-fitted models to predict the target variable, i.e., total thermal load, given the input features.

### 2.1 Data Generation

A hypothetical open office plan is parametrically generated in Rhino/Grasshopper considering. The energy consumption of all building models are simulated parametrically through Honeybee with the EnergyPlus engine considering 100 m$^2$ area with 5 meter height (minimum length of building side is 2.5 m). The four building outlines selected for this study are displayed in Figure 1.

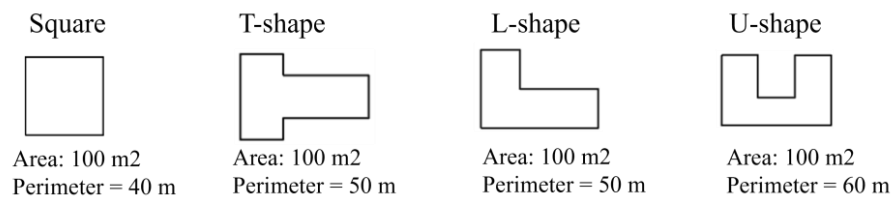

**Figure 1.** Building plan outlines based on (Hatem and Karram 2020)

This study has used a statistical method to consider uncertainty in the simulation process to address the deviations between the pre-assumed/simulation outcomes and buildings' performances in practice reported in literature (Bae *et al.*, 2020, De Wit 2001). Artificial noise was employed to impose uncertainty to the simulated values of the features to calculate the buildings' energy consumptions with probable deviations that occur in practice. A common choice of artificial noise is a random variable distributed as a univariate zero-mean Gaussian distribution. In other words, for the j$^{th}$ feature a noise can be sampled from a univariate zero-mean Gaussian distribution with a standard deviation of σ$_j$. In this study, for each feature, a set of possible values are considered. Tables 1 and 2 represent all variables (building features) utilized in this study. Table 1-column 2 shows the possible values considered for each feature and Table 1-column 3 represents the number of possible values associated with the feature. The total combination of all features leads to 1440 (5×4×3×12×2) conditions per building shape. Hence, in this study, we generated 5760 samples (1440 per building outline). The amount of means and standard deviations of the variables corresponding to the wall material specifications, i.e., wall thickness, conductivity, density, and specific heat capacity, have been derived from studies in literature (Macdonald 2002, Hopfe 2009) (Table 2). For the rest of the features, due to the lack of literature, we added a Gaussian noise with σ = 0.01 per each possible value (Table 1-column 4). We determined this amount to generate a normal distribution around each possible feature value while respecting the intervals between them so that they don't overlap. However, we considered a larger variance for the building orientation parameter to address the high uncertainty of buildings' orientations during the early design stages (Macdonald 2002). Studies show that a perturbation of 5° on the building orientation could result in a significant uncertainty on the solar fraction that is often overlooked by designers (Almeida Rocha *et al.* 2016). Hence, to address the high level of uncertainty and cover a wide range of buildings' orientations due to the variation of actual urban grid orientation, we used σ = 3 for the building orientation parameter. This amount will cover the perturbation of 5° reported in literature in a probabilistic approach. Other energy simulation parameters have been considered constant across all samples. We adopted those values from ASHRAE standards (ASHRAE Standards, 2013) used in similar studies (John Haymaker *et al.* 2018).

Table 1. Buildings' variables, sets of possible values, and standard deviations used in data sampling

| Variables | Set of Possible Values | Number of Possible Values | Standard Deviation (σ) |
|---|---|---|---|
| WWR | {0.1, 0.2, 0.3, 0.4, 0.5} | 5 | 0.01 |
| Windows' shading depth (m) | {0, 0.15, 0.30, 0.45} | 4 | 0.01 |
| Glazing U-value (W/(m2·K)) | {0.7, 2.72, 4.54} | 3 | 0.01 |
| Building Orientation (degree) | {0º, 30º, 60º, …, 330º} | 12 | 3 |
| External Wall Material * | * Table 2 | 2 | * Table 2 |

Table 2. External wall material choices along with means (μ) and standard deviations (σ) adopted from literature

| * External Wall Material | Thickness (m) | Thermal Conductivity (W/m·K) | Density (kg/m³) | Specific Heat Capacity (J/kg·K) |
|---|---|---|---|---|
| Concrete | (μ =0.21, σ =0.021) | (μ =1.13, σ =0.1) | (μ =2000, σ =30) | (μ =1000, σ =106) |
| Brick | (μ =0.16, σ =0.016) | (μ =0.84, σ =0.27) | (μ =1700, σ =297.5) | (μ =800, σ =86) |

To generate the data samples the following method is applied:

Let ($x_n$, $y_n$) be the $n^{th}$ sample, where $x_n$ is an 8-dimensional feature vector and $y_n$ is the corresponding response variable (i.e. building's energy consumption). Each data point is generated as described in the pseudo code below (Listing 1), and is used as an input vector to the energy simulator to generate the corresponding y value.

**Listing 1** Pseudo code to generate the $n^{th}$ data sample

1.  For j in [1, 2, …, 8]:
      Randomly, select a value from the set of possible values corresponding to feature j
      Add a Gaussian noise with $σ_j$ around the selected value
2.  Generate the $n^{th}$ sample $x_n$ = [$x_1$, $x_2$, … ,$x_8$], where $x_j$ is the $j^{th}$ feature

3     Generate $y_n$, by sending $\mathbf{x}_n$ to the energy simulator
4     Return ($\mathbf{x}_n$, $y_n$)

## 2.2 Data Analysis

To analyse the data, multiple methods have been utilized. Initially, we adopted exploratory data analysis to assess the impact of building's shape on the total energy consumption rate. Figure 2 depicts the probability distributions of thermal loads corresponding to the four selected outlines.

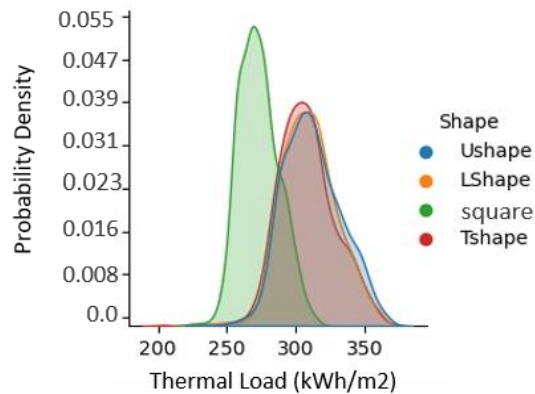

**Figure 2.** Probability distribution plot of thermal loads across the 4 selected buildings' shapes

Figure 2 demonstrates that the thermal loads of building models with T-, U-, and L-shapes have similar probability distributions although they have different wall-to-floor ratios (see Figure 1). Also, the average thermal load of buildings with square outlines (272.92 $kWh/m^2$) is lower than the average thermal loads of buildings with T-,U-, and L-shapes (309.64 $kWh/m^2$). Furthermore, the standard deviation of the thermal loads corresponding to the square shape (14.53) is lower than the other three shapes (20.4 on average). In other words, considering noisy features that commonly occur in practice, thermal loads of buildings with square shapes are mostly distributed around the mean; hence, the simulated values tend to be closer to the expected values with higher probability compared to the other outlines. Expected values refer to the thermal loads calculated with deterministic features that are commonly used in energy simulation methods. Table 3 represents the quantitative comparative analysis of the building models' thermal loads with respect to their shape outlines.

Table 3. Minimum, maximum, mean and standard deviation pertained to thermal loads of building models with different shape outlines

| Building Shape | Minimum Thermal Load (kWh/m2) | Maximum Thermal Load (kWh/m2) | Mean Thermal Load (kWh/m2) | Standard Deviation (σ) |
|---|---|---|---|---|
| Square | 222.27 | 317.16 | 272.92 | 14.53 |
| T-shape | 203.54 | 368.61 | 307.63 | 20.53 |
| U-shape | 234.18 | 369.54 | 311.56 | 20.9 |
| L-shape | 232.34 | 362.64 | 309.72 | 19.74 |

Table 3 leads to the following conclusions:

- The average minimum thermal loads of building models with T-, U-, and L-shapes is 0.5% higher than the minimum thermal load of building models with square shapes.

- The average maximum thermal loads of building models with T-, U-, and L-shape is 13.6% higher than the maximum load of building models with square shapes.

- The average thermal load of building models with T-, U-, and L-shape is 11.9% higher than the average thermal load of building models with a square shapes.

- The average deviation of the thermal loads of building models with T-, U-, and L-shape is 28.98% higher than the deviation of the thermal load of building models with square shapes from their corresponding means.

It is important to mention that the results reflect the significance of building outlines besides the wall-to-floor ratio impact on buildings energy consumptions.

### 2.2.1 Variables' correlations and significance

Principal Component Analysis (PCA) is a non-parametric statistical technique for data dimensionality reduction (Wold et al. 1987, Ghane et al. 2015). PCA transforms data points along different axes which contain information of the original variables. The transformation process (projection) produces linear combinations of the original variables with the eigenvector's coefficients through axes known as PCs (Principal Components) (Ghane 2015). Hence, the total number of PCs aligns with the number of variables although the PCs do not coincide exactly with any of the original variables. Through PCA we can assess the relationship among variables and their impact on data distribution. Depending on the variables' correlations, often the first few PCs explain the major variation of data. The results from PCA could be utilized to summarize and visualize the informational content of a large data with multiple variables through tables and figures. The loading matrix derived from the PCA method shows the weight, i.e., coefficient, of each variable across the corresponding PC. Hence, the absolute value of the weights explain how much each variable contributes to a specific PC. The negative or positive sign only represents the direction of the corresponding eigenvector. Thus, large values, regardless of their sign, indicate a strong contribution of a variable to a particular PC. Since the primary PCs represent the major variation of data, the weights of the corresponding variables indicate their significance to data distribution. Figure 3 depicts the scree plots of the PCs; left: percentage of variance explained by each PC, and right: the cumulative percentage of the variance retained by the PCs.

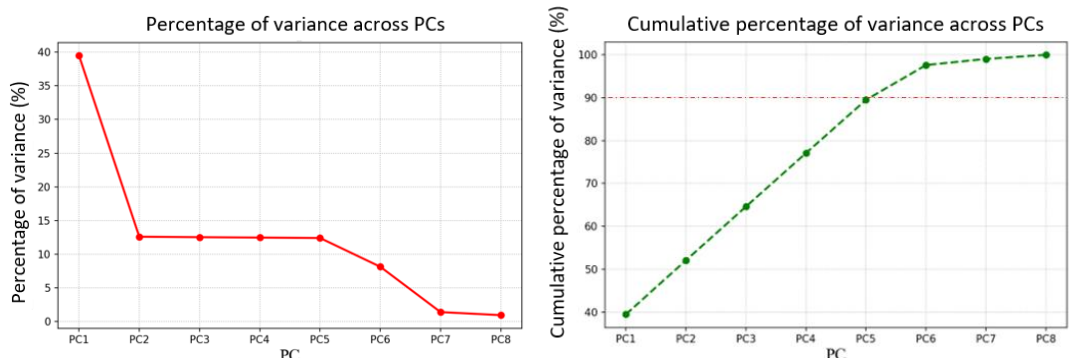

**Figure 3.** Left: Percentage of variance of PCs; Right: Cumulative percentage of variance of PCs

Figure 3-left demonstrates that PC1 explains ~40% of data variation, while the next four PCs (PC2 to PC5) account for ~12.5% of data variation each. Based on the cumulative (sum) of the variance percentages, the primary five PCs are required to explain 90% of the variance in the data (Figure 3-right).

Table 4 shows the loading matrix, including the variables and their corresponding coefficients across each PC. The highlighted cells indicate the variables' weights which have relatively strong contributions to the corresponding PCs among PC1-PC5.

Table 4. Loading matrix; coefficients of variables across PCs

| Variables | PC1 | PC2 | PC3 | PC4 | PC5 | PC6 | PC7 | PC8 |
|---|---|---|---|---|---|---|---|---|
| Orientation | 0.0066356 | 0.2367618 | 0.759922 | 0.23574 | -0.55666 | -0.01739 | -0.02938 | 0.001466 |
| WWR | -0.001685 | -0.138909 | 0.540427 | -0.74829 | 0.357905 | 0.01706 | -0.02218 | -0.00498 |
| Shading depth | 0.0194056 | -0.682932 | 0.320954 | 0.534051 | 0.373885 | -0.07206 | 0.009134 | -0.01162 |
| Glazing U-value | 0.0140875 | -0.679363 | -0.17096 | -0.30661 | -0.64403 | 0.004916 | -0.01373 | -0.01547 |
| Wall thickness | -0.832547 | -0.029016 | -0.00622 | 0.003005 | 0.000642 | -0.04546 | -0.20402 | 0.512328 |
| Wall conductivity | -0.698691 | -0.039503 | 0.044839 | 0.045921 | 0.000365 | 0.629432 | 0.323426 | -0.07456 |
| Wall density | -0.719927 | 0.02969 | 0.002896 | -0.05547 | -0.02847 | -0.51854 | 0.443622 | -0.10698 |
| Wall SHC | -0.797497 | 0.0117432 | -0.02544 | 0.017823 | 0.017048 | -0.03774 | -0.47105 | -0.37348 |

Table 4 displays that wall material specifications with relatively high coefficients significantly contribute to PC1. The rest of the variables, i.e., orientation, WWR, shading, and glazing U-value, also contribute to PC2-PC5 with relatively high coefficients. As previously mentioned, PC1 and PC2-PC5 explain 40% and 50% of the data variation, respectively. Therefore, all features selected in this study have relatively high or nontrivial correlation across at least one of the primary PCs; hence, no feature is eliminated for training the prediction models in the next section.

Note that PCA is an unsupervised method meaning that it does not take into account the response variable. The PCs are orthogonal axes representing the linear functions of the variables with coefficients corresponding to each variable; however, since PCA does not consider the response variable, i.e., thermal load, the results cannot be inferred as a sensitivity analysis of the features. Saying that, summary statistics (e.g., min, max, mean, etc.) may not be conducted upon PCA loading matrices. In this study PCA was only used as a feature selection technique to determine whether any of the selected dimensions (variables) could be neglected.

### 2.3 Prediction Models

Finally, the study has explored Polynomial regression models with different degrees of freedom (degree 1 to 4) to predict the response variable, i.e., thermal load, knowing the input features. The degrees in the polynomial models affect the complexity of the models. For example, the simplest polynomial model is a linear regression model with a degree of 1. Note that regression models require numerical values for all features. Therefore, we had to either ignore the form feature or consider separate models for each form. Assigning dummy variables with numerical values to the form feature may not be an appropriate solution since each form is not numerically better or worse than the other one. Hence, two conditions are studied, including: 1) using all data ignoring building-shape feature and 2) splitting data based on building-shape feature and using separate models for each case. Based on our primary analysis explained in the "Data Analysis" section, the data is split into two sets of samples in the second condition, including building models with square shapes and building models with T-, U-, and L-shapes. R squared ($R^2$), i.e., coefficient of determination of the prediction model, is used as a metric to determine the goodness of the fitted model. $R^2$ normally ranges between 0 and 1 (0% - 100%) where 0 is the worse and 1 means the exact fit. Another issue to address in this study is overfitting phenomena. Overfitting is a common incident that happens when the model fits too much into the training data and fails in predicting the test set in the real world (Tom Dietterich 1995). Different methods are introduced in literature to avoid overfitting (Bartlett *et al.* 2020, Braga-Neto 2020). In this study, although statistically we have a huge amount of data, all data is synthetically generated from one source; hence, train and test samples are very similar to each other. To generalize the model and avoid overfitting, only a small portion of the data (30%) is used for training, and the rest is kept for testing (70%). Table 5 shows the results of $R^2$ on test data using polynomial regression models with different degrees across the two conditions as follows:

1) Considering all samples and ignoring building-shape feature (Table 5, condition 1).

2) Samples of building models with square outline versus T-, U- and L-outlines; hence, considering separate models for each of the two categories (Table 5, condition 2).

The training time corresponding to each predictive model is also calculated (column-3) to reveal the impact of model complexity on computational time.

Table 5. R-squared and training time pertained to prediction models considering the two conditions

| | | Prediction Model | $R^2$ (%) | Training Time (milliseconds) |
|---|---|---|---|---|
| Condition 1 | All samples ignoring form feature | Linear Regression (degree = 1) | 71 | 0.0047 |
| | | Polynomial Regression (degree = 2) | 79.5 | 0.0083 |
| | | Polynomial Regression (degree = 3) | 78.7 | 0.0785 |
| | | Polynomial Regression (degree = 4) | 74.3 | 233.27 |
| Condition 2 | Samples with square shapes | Linear Regression (degree = 1) | 86.9 | 0.0017 |
| | | Polynomial Regression (degree = 2) | 98.3 | 0.0064 |
| | | Polynomial Regression (degree = 3) | 97.7 | 0.069 |
| | | Polynomial Regression (degree = 4) | 88 | 358.2476 |
| | Samples with L, T, and U shapes | Linear Regression (degree = 1) | 87.2 | 0.003 |
| | | Polynomial Regression (degree = 2) | 98.4 | 0.008 |
| | | Polynomial Regression (degree = 3) | 98.6 | 0.0803 |
| | | Polynomial Regression (degree = 4) | 98.3 | 244.68 |

The results show that considering separate prediction models for square shape buildings versus buildings with T-, U-, and L-shapes can better fit the data. Considering one general model in predicting the response variable of all buildings with different shapes has not exceed $R^2$ of 79.5%, whereas the separate prediction models respecting building shape feature could reach to $R^2$ of 98%. This result matches with the analysis explained in the "Data Analysis" section, validating the difference between thermal loads' distribution of buildings with square outline versus T-, U-, and L-outlines. Therefore, the results confirm the impact of buildings' shapes on total energy consumptions. Also, in most cases, the polynomial models with degree of 2 are the best fitted models. More complex models with higher degrees have not necessarily resulted in better fits for this dataset while they take exponentially more computational time to get trained. The computational time is calculated based on the current study training sample size and could increase if the training input increases. The prediction model for T-, U-, and L-shapes shows better result with degree = 3 with only 0.2 % improvement compared to the model with degree = 2; the computational time has increased by 90% which is a significant leap.

The plots in Figures 4 to 6 display the relation between predicted values to simulated values of test data using the selected prediction models over the samples in the two conditions previously explained. The degrees of the models have been increased form 1 to 4 (left to right) in the figures where the most left plots correspond to the models with degree = 1 (Linear Regression), and the most right plots correspond to the polynomial models with degree = 4.

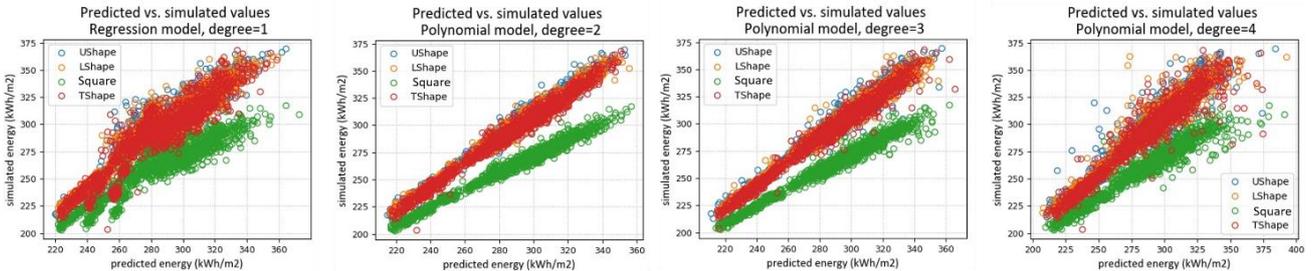

**Figure 4.** Plots of polynomial models (degree 1 to 4, from left to right) considering all data, ignoring building outline feature.

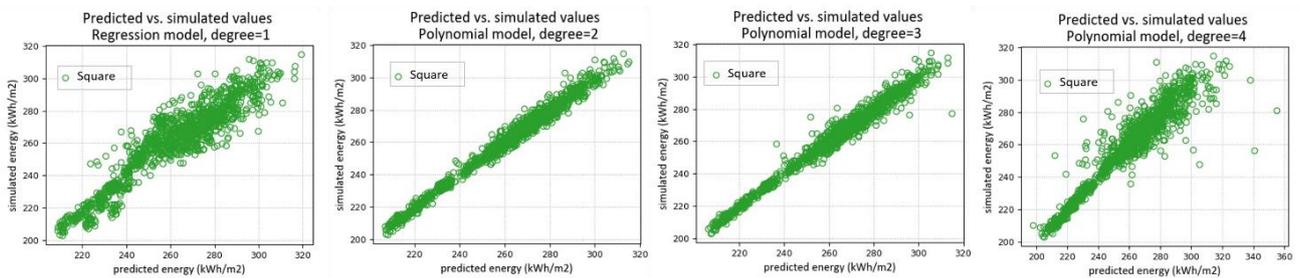

**Figure 5.** Plots of polynomial prediction models (degree 1 to 4, from left to right) for samples of square-outline building models.

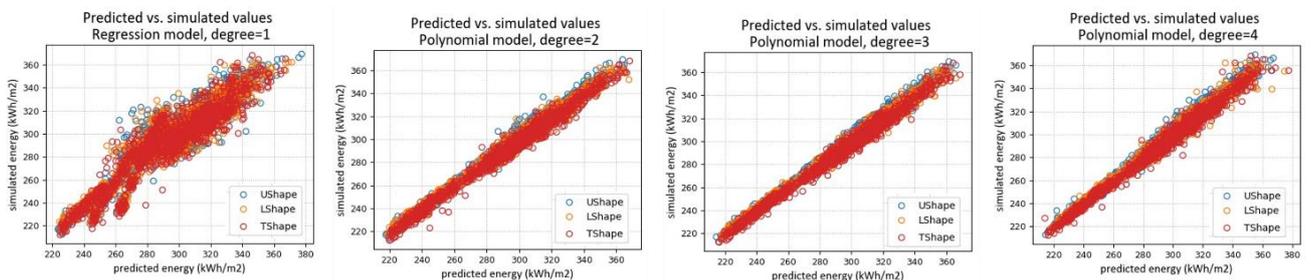

**Figure 6.** plots of polynomial prediction models (degree 1 to 4, from left to right) for samples of T-,U-, and L-shape building models.

## 3   Conclusions and Further Research

This study explored the impact of buildings' outlines on developing accurate predictive models to predict buildings energy consumptions. Eight performance-based parameters are considered in generating synthetic data for each building outline. Due to the deviations reported in literature between simulated and actual values, we implemented Gaussian noise to apply probability to the features values. In total, 5760 feature vectors were synthesized (1440 per building outline). Subsequently, the feature vectors were used to simulate the thermal load.
Our results show that the probability distributions of the simulated thermal loads were similar for buildings with T-, U-, and L-shape outlines although they had different wall-to-floor ratios. The buildings with square outlines, on average, consumes 11.9% less thermal loads than buildings with T-, U-, and L-shape outlines. This result matches with the study done by Hatem et al., 2020, confirming that square outlines show more efficiency in energy consumption compared to other outlines with similar area (Hatem and Karram 2020). The results also reflect that considering uncertainty for feature values, thermal loads of buildings with square shapes are mostly distributed around the mean with a higher probability (mean = 272.92 , std = 14.53) compared to the other shapes (average mean = 309.63 , average std = 20.4). Therefore, we may expect the thermal loads of the square-outline building models with lower uncertainty around the mean compared to other shapes.

Additionally, we investigated the contribution of features in the data variation by means of PCA. The result form PCA method indicates that PC1 retains 40% of data variation, whereas the set of next four PCs explain 50% of the variation. Hence, the primary 5 PCs are required to explain ~90% of data variation. The major contributors to PC1 are wall material specifications, i.e., wall thickness, conductivity, density and specific heat capacity. For the next four PCs, other performance-based parameters, i.e., building orientation, WWR, glazing U-value, and window shading depth, have significant impact. Therefore, all of the aforementioned features are important in energy consumption of the synthetic data and may not be neglected for training predictive models.

Based on the analysis of the impact of a building outline on the response variable, i.e., thermal load, predictive regression models with different complexities are developed. The regression models with degree of 2 perfectly fit the data in most cases while being computationally and timely more efficient than models with higher degrees. In addition, models with higher degrees are prone to overfit the training data, and therefore may not predict the unseen samples with high accuracy. We assessed the goodness of fits using R squared metric. The results show that fitting a mono-predictive model for all samples, regardless of outlines yielded $R^2$ = 79.5 %. The goodness of fits increased by fitting two separate models; one model on the samples of square shape ($R^2$ = 98.3%) and one model on samples with the T-, U-, and L-shapes ($R^2$ = 98.6%).

In the future study, the authors intend to consider other building outlines and their variations to cover vast variety of building shapes. Also, we plan to develop more complex non-linear algorithms, such as SVR (Support Vector Regression) and DNN (Deep Neural Network), to examine if a single, complex model could accurately fit all data variations, yet not overfit the training data, covering several building outlines.